\documentclass[aps,prd,superscriptaddress,floatfix]{revtex4-1}
\usepackage{color}
\usepackage{amssymb}
\usepackage[normalem]{ulem}
\usepackage{graphicx,color}
\usepackage{amsmath}
\usepackage{subfigure}
\usepackage[normalem]{ulem}
\usepackage{booktabs}
\usepackage{xcolor}
\usepackage{epsfig,graphics,color,graphicx,amsmath}

\colorlet{darkgreen}{green!50!black}
\colorlet{brightyellow}{yellow!75!red}
\colorlet{orange}{red!50!yellow}
\colorlet{darkblue}{blue!60!black}
\colorlet{darkred}{red!80!black}

\usepackage{soul}

\usepackage{mathtools}
\usepackage{mathrsfs}
\usepackage{bm}
\usepackage{relsize}
\usepackage{indentfirst}

\usepackage{xcolor}

\usepackage{amssymb,amsmath,multirow,epsfig,graphicx,color}

\usepackage{epsfig}
\usepackage{graphicx,amsmath}
\def\be{\begin{eqnarray} &&}

	\def\ee{\end{eqnarray}}

\newcommand\ba{\begin{eqnarray}}
	\newcommand\ea{\end{eqnarray}}

\newcommand{\bas}{\begin{eqnarray*}}
	\newcommand{\eas}{\end{eqnarray*}}

\newcommand{\bno}{\begin{eqnarray*}}
	\newcommand{\eno}{\end{eqnarray*}}

\def\sl
\usepackage{hyperref}
\usepackage{url}
\usepackage{hyperref}
\hypersetup{
	colorlinks=true,  
	linkcolor=blue,   
	filecolor=magenta,      
	urlcolor=blue,    
	citecolor=blue,   
} 
\begin{document}
	\vspace{-12ex}    
	\begin{flushright} 	
		{\normalsize \bf \hspace{50ex}}
	\end{flushright}
	\vspace{11ex}
	\title{Quasi-de Sitter Solution, Negative Norm State and Krein Space in Cosmology}
	\author{M. Mohsenzadeh}
	\email{ma.mohsenzadeh@iau.ac.ir}
	\affiliation{Department of Physics, Faculty of Basic Sciences, Qom Branch, Islamic Azad University, Qom, Iran}

	\date{\today}
	
	\begin{abstract}
		Recent observational data have shown two important results about cosmic inflation: First, the primordial power spectrum is approximately scale invariant. Secondly, the space-time geometry during inflation is quasi-de Sitter. In this paper, we study the background field method and the concept of negative norm states in the framework of Krein space to use in early universe cosmology. For the inflaton field equation, based on the observations and taking into account our studies, we suggest an alternative proposal. We will  take the quasi-de Sitter solutions as the physical geometry with the positive norm and the pure de Sitter solutions as the background geometry with the negative norm. In the following, we review the consequences of applying the proposal to calculate some cosmological quantities.\\
		\noindent \hspace{0.35cm} 
		\noindent \hspace{0.35cm} 
	\end{abstract}

	\maketitle
	
\section{Introduction}
Describing gravity using Einstein’s general relativity equation does indeed lead us to the theory of quantum fields in curved space-time. In the absence of complete and efficient theory of quantum gravity, it is not possible to investigate the quantum phenomena that result from attendance a gravitational field. At the beginning of the emergence of the quantum theory, in the calculation of the phenomenon involving electromagnetic interaction between elementary particles, the electromagnetic field were considered as a background classical field, which interacts with the quantum matter. This was a semi-classical method, and the results of this method was fully compatible with the theory of quantum electrodynamics. Thus, it is expected that a similar approach regarding gravity, in which the gravitational field is the sum of a classical background field and a material field that is quantum, can achieve the effects of quantum phenomena with the presence of the gravitational field.\\ At the beginning of the emergence of the quantum theory, in the calculation of the phenomenon involving electromagnetic interaction between elementary particles, the electromagnetic field were considered as a background classical field, which interacts with the quantum matter. This was a semi-classical method, and the results of this method was fully compatible with the theory of quantum electrodynamics. Thus, it is expected that a similar approach in the case of gravity, in which the gravitational field is the sum of a classical background field and a material field that is quantum, can achieve the effects of quantum phenomena with the presence of the gravitational field.\\
The space-time in general relativity is considered to be curved, and as a result, in quantization of space-time (ie, the quantization of $g_{\rm \mu\nu}$ as a gravitational field) calls into question causality. Because $g_{\rm \mu\nu}$ determine light cone, and when it considered as an operator, it means that the light cone fluctuate around the mean value, and this oscillation causes a lack of clear definition of causality. This problem was solved by introducing the background field by De Witte, in which the metric is divided into two parts \cite{DeWitt:1967Jjm,Birrell:1982ip}:
\begin{equation}
	\label{f1}
	\hat{g}_{\rm \mu\nu}=\bar{g}_{\rm \mu\nu}+\hat{h}_{\rm \mu\nu},  
\end{equation}                     
where $\bar{g}_{\rm \mu\nu}$ is a background metric and is considered to be classical, and $\hat{h}_{\rm \mu\nu}$ is metric perturbation and is considered as a quantum operator. In this case, the principle of causality is explained according to the background metric $\bar{g}_{\rm \mu\nu}$, which is non-operator, and $\hat{h}_{\rm \mu\nu}$ which is propagated in the background metric, can be quantized. However, this theory is not generally renormalizable, although the quantization in one loop is renormalizable. It should be noted that in order to have covariant quantization, the background metric must be curved \cite{Isham:1974ip}.
\par After relativistic quantum mechanics, quantum field theory was an important achievement that scientists in theoretical physics used to describe the physical phenomena that had previously been obscure to humans. But this theory as well as previous theories have a fundamental problem. This big problem is the inability to create unification among all forces, and of course, it has not yet been achieved. In fact, this theory is not able to quantize gravity. The emergence of divergence in quantum gravity and the inability to eliminate these divergences by existing renormalizable theories have led to the failure of quantum field theory to arrive at a complete structure that describes all physical events.\\ One way to achieve a complete quantum theory is to use of the generalized Hilbert space. In this space, we use of the solutions whose norm is negative\cite{Takook:2000zj,Takook:2023bsb}. One important result of using this space in the quantization process, s covariant quantizations of the massless scalar field with a coupled minimum in de Sitter space-time \cite{Dehghani:2008zza}.
\par Data from the Wilkinson probe and Planck satellite on cosmic background microwave radiation indicate that, with approximately 95$\%$ confidence, the scalar spectral index value is within the range of $n_s=0.9649 \pm 0.0042$ \cite{Planck:2018jri}. Therefore, cosmologists believe that the universe, in the period of inflation, is similar to de Sitter in the first approximation \cite{Yusofi:2018lqb,Suratgar:2022tsi,Takook:2023bsb,Geshnizjani:2023edw}. Since the theory of quantum fields is very closely related to early universe cosmology, it seems natural that in this branch of physics we be also seek to examine the results of new methods adopted in quantization of the fields. On the other hand, given that the theory of quantum fields in curved space plays a significant role in the discussion of inflation, so it can be expected that the result of considering the period of quasi-de Sitter inflation can affect the initial value of the quantum fluctuations of the scalar field in the early universe \cite{Kempf:2001fa}. In fact, these effects can be reflected in the power spectrum of temperature fluctuations in cosmic background radiation \cite{Mohsenzadeh:2008yv,Mohsenzadeh:2013bba,Mohsenzadeh:2014zfa, Mohsenzadeh:2016pxd,Yusofi:2014mma,Yusofi:2014mta,Yusofi:2018lqb,Ziyaee:2021pno}, which today, with the help of observational information sent to us by satellites, it has provided an important tool for judging the theories and models of cosmologists \cite{Suratgar:2022tsi}. Reviewing of the published paper by using our proposal, it will be shown that the choice of non-flat background geometry will have an impact on the final results. It will also be more appropriate to use a non-flat background in cosmology because the general convergence principle in the Krein method for cosmology can become establish \cite{Ziyaee:2020wik,Ziyaee:2021pno}. For this purpose, in the section \ref{Sec 2.}, by considering the background field method, we introduce the quasi-de Sitter space-time and it's relation with Krein space. In section \ref{Sec 3.}, we use quasi-de Sitter solutions and the Krein space to calculate cosmological spectrum. We review results of our calculations by selection different physical geometry in our covariant method in \ref{Sec 4.} section. In section \ref{Sec 5.}, concluding remarks are presented.
\section{Quasi-de Sitter Geometry and Krein Space} 
\label{Sec 2.}
For several reasons, in recent years, the attention of physicists, especially cosmologists, has been drawn to the pure de Sitter space-time. Among these reasons, apart from the maximum symmetry, is the very simple geometry of this space-time, which makes it a great experimental model for studying the physics of the universe. Also, in terms of observations, cosmic inflation is a quasi-exponential expansion that results from solving de Sitter of Friedmann's equations.\\
In the standard view, based on the background field method (\ref{f1}), we can write for the geometry of the universe \cite{Mohsenzadeh:2008yv}:
\begin{equation}
	\label{f2}
	\hat{g}_{\rm \mu\nu}= \bar{g}^{\rm min}_{\rm \mu\nu}+\hat{h}_{\rm \mu\nu}  
\end{equation}                     where $\hat{h}_{\rm \mu\nu}$ is a perturbation on the Minkowski background $\bar{\eta}_{\rm \mu\nu}$. In this view, the total geometry $\hat{g}_{\rm \mu\nu}$ is generally taken to be pure de Sitter, and the background geometry is Minkowski $\bar{g}^{\rm min}_{\rm \mu\nu} = \bar{\eta}_{\rm \mu\nu}$. Therefore, from the perspective of quantum field theory, the physical result of this work is like to renormalization of quantum field theory in curved space-time relative to the flat Minkowski space-time.\\
As mentioned in the introduction, the observations show that in the inflationary period, space-time is not pure de Sitter, but in the approximation of the first order, it is pure de Sitter. So with this view and based on the background field method (\ref{f1}), we can write for the geometry of the universe \cite{Mohsenzadeh:2008yv}:
\begin{equation}
	\label{f3}
	\hat{g}_{\rm \mu\nu}= \bar{g}^{\rm dS}_{\rm \mu\nu}+\hat{h}_{\rm \mu\nu}  
\end{equation}                 
where $\hat{h}_{\rm \mu\nu}$ is a perturbation on the pure de Sitter background $\bar{g}^{\rm dS}_{\rm \mu\nu}$. In this view, the total geometry is usually taken to be quasi-de Sitter (asymptotic or approximation), and the pure de Sitter is background geometry. Therefore, from the perspective of quantum field theory, the physical result of this work is like to nonstandard renormalization of quantum field theory in curved space-time relative to the pure de Sitter space-time. Given that the geometry of the total universe is not pure de Sitter, but in the first approximation is pure de Sitter, therefor if we do the calculations using method (2), in fact we have just reached the first approximation of the geometry of the universe and not the actual geometry of the total universe. If we do the calculations using method (3), although we may not have reached the true geometry of the total universe, but the result is closer to reality than method (2).\\
According to the perturbation theory, the perturbation sentence in (\ref{f1}) must be much smaller than the background, so that the theory is valid. However, considering that the geometry of the total universe has a non-zero curvature, then in the expression (\ref{f2}), since the background sentence does not have a curvature, so the perturbation sentence must have a non-zero curvature. Thus, this causes the perturbation sentence become larger than the background, and this undermines the validity of the perturbation theory. But in the expression (3), in order to validate the perturbation theory, it can be assumed that the amount of curvature of the perturbation sentence is much smaller than the amount of curvature of the background sentence.\\
Perhaps one of the reasons that Minkowski space-time is used in method (2) as a background geometry is that its curvature is a constant value (equal to zero). If this is one of the criteria for selecting suggested that in method (3) pure de Sitter space-time can also be selected as background geometry, because its curvature has a constant value ($12H^2$). Perhaps another reason for choosing the Minkowski space-time as background geometry is that it has maximum symmetry. So since pure de Sitter space-time also has maximum symmetry, it can also be selected as background geometry.\\
For pure de Sitter space-time, the symmetry group is different from the symmetry group of Minkowski space-time. So, another reason for the choice of pure de Sitter space-time as background geometry is that the covariance of symmetry group of the total geometry in method (\ref{f3}) is maintained, while with the choice of Minkowski space-time as geometry background the covariance of symmetry group of the total geometry in method (\ref{f2}) is not maintained. Also, it seams that the total geometry is more similar to pure de Sitter space-time than to Minkowski space-time, because the curvature of the total geometry and the pure de Sitter is non-zero, while the curvature of Minkowski is zero.

\subsection{Use of Krein space in cosmology}
Krein space is a vector space in which a seemingly basic condition of the vector space is ignored. This basic condition is that the norm of vectors being positive. Such spaces, in which the inner product has both positive and negative signs, are known as the Krein space \cite{Gazeau:1999mi}. In other words, the norm in this space is indefinite. Therefore, according to the definition of Krein space, a Hilbert space can be combined with an anti-Hilbert space, whose members are the complex conjugate of Hilbert space, and so a space with more freedom of action can be achieved. We can have two interpretations for each of Hilbert and anti-Hilbert spaces. For Hilbert space can say: (A) The space in which the norm of states is positive (from the perspective of mathematicians), and (B) The space that is related to the physical world (from the perspective of physicists).\\ Also for anti-Hilbert space can say: (A) The space in which the norm of states is negative (from the perspective of mathematicians), and (B) The space that is related to the non-physical world (from the perspective of physicists).
Another point that can be made with respect to the above discussion is the choice of solutions according to the definition of the horizon in cosmology. From a cosmological point of view, we can choose the solutions outside the horizon as non-physical solutions without their norm being negative. We can also choose the solution within the horizon as a non-physical solution if we have a negative norm\cite{Mohsenzadeh:2008yv,Mohsenzadeh:2011zz,Mohsenzadeh:2013bba}.\\
As we know, through imposition a physical interaction on a field operator, only states with positive norm would be affected. The negative norm States do not interact with physical states (or real physical world), so they are not impressed by physical interactions. These negative norm states are similar to the ghost states in quantum field theory. In gauge quantum field theory, the ghost states cannot be propagated in the physical world, nor can interact with physical states. In the discussion of quantum fields, several articles have shown that the states with negative norm can play the role of a renormalization tool \cite{Takook:2000zj,Rouhani:2004zs,Khosravi:2006tx}.\\
On the other hand, one thing that is important in terms of mathematical structure for choosing positive and negative solutions in Krein space, and it should be emphasized, firstly, is that these solutions must be from two seemingly different spaces (one called Hilbert space and one called anti-Hilbert space) and secondly, the properties of the solutions (for example, sign behind them) must be somewhat different. According to the mathematical definition, Krein space is made up of the direct summation of Hilbert and anti-Hilbert space. But the fact that the Hilbert and anti-Hilbert spaces make up the Krein space does not necessarily mean that the negative solution must be a complex conjugate of the positive solution. Therefore, according to the background field method (\ref{f1}), the positive solution can be selected from the modes related to an original (physical) curved space-time, and the negative solution can be selected from the modes related to the background curved space-time.\\
To make the above statement more complete, we will explain the discussion with an example. In the discussion of parity, we know that in classical physics, parity is applied to the variable itself, such as x. But in quantum physics, the parity does not apply to the variable x itself, but to the expected value of the variable x. On the other hand, we know that in the use of Krein space in QFT (and in flat space), the positive and negative sign is exerted to the norms. In using Krein space in cosmology (where curved space is commonly used), we suggest that positive and negative sign be exerted to the expected value (or two-point function). That is, just as in QFT (and in flat space) we choose a set of solutions with a positive norm and a set of solutions with a negative norm, so in cosmology we choose a set of solutions that give us a positive expected value (or two-point function), and we choose another set of solutions that give us a negative expected value (or two-point function).\\
The reason for emphasizing the use of expected value (or two-point function) instead of using the norm is that, in a flat space-time, the norm is definable, so we can talk about its sign. But in curved space-time, the norm is indefinable, so we can not talk about the sign of the norm. So, as in cosmology we usually use curved space-time, then naturally the norm in cosmology can not be defined. For this reason, we suggest that in cosmology, instead of using the norm, we use the expected value (or two-point function). So, the sign of the expected value (or two-point function) in cosmology is equivalent to the sign of the norm in flat space-time. Therefore, it is suggested to attribute the positive and negative sign related to the expected value (or two-point function) to the sign behind the phrase related to the expected value (or two-point function).
\begin{equation}
	\label{f4}
	\langle A \rangle-\langle B \rangle=\langle A \rangle +(-\langle B \rangle)
\end{equation}                     
where, according to the symbols behind the sentences in the phrase on the right, it can be said that the symbol behind the first sentence represents the expected value (or two-point function) positively and the sign behind the second sentence represents the expected value (or two-point function) negative. The standard method in quantum field theory in Minkowski space-time for calculating finite value of the energy is using normal ordering to eliminate the amount of ultraviolet divergence associated with the vacuum energy of the quantum field. The standard method in quantum field theory in curved space-time for the renormalization of the ultraviolet divergence of vacuum energy is the subtracting local divergencies of Minkowski space-time \cite{Birrell:1982ip}:
\begin{equation}
	\label{f5}
	\langle \Omega|:T_{\rm \mu\nu}:|\Omega \rangle= \langle \Omega|T_{\rm \mu\nu}|\Omega \rangle -\langle 0|T_{\rm \mu\nu}|0 \rangle =  \langle \Omega|T_{\rm \mu\nu}|\Omega \rangle +(-\langle 0|T_{\rm \mu\nu}|0 \rangle)
\end{equation}                                        
where $|\Omega \rangle$ is the initial state in curved space-time and $|0 \rangle$ is the initial state in Minkowski space-time. Here divergences are defined in Minkowski space-time. In fact, they are not the solutions to the wave equation in curved space-time. Therefore, this renormalization clearly breaks the covariance of curved spacetime. But there is another method to remove local divergences in curved space-time (called the non-standard renormalization method). In this method, local divergences of curved space-time are removed by quantities defined in selfsame curved space-time. In fact, they are the solutions of the wave equation in curved spacetime. This procedure of renormalization seems more logical because the covariance of curved space-time is preserved. It is necessary to mention that the problem of the ”standard” renormalization method is that the initial states are defined non-locally, while divergences are removed locally.
\section{Application of quasi-de Sitter geometry and Krein space in cosmology}
\label{Sec 3.}
The metric below is commonly used to define the universe during inflation\cite{Baumann:2009ds}:

\begin{equation}
	\label{f6}
{ds}^{2} = {dt}^{2} - a^{2}(t){d\overrightarrow{x}}^{2} = {a(\tau)}^{2}({d\tau}^{2} - {d\overrightarrow{x}}^{2})
\end{equation}

where the scale factor $ a(\tau)$ is the function of the
conformal time $\tau$. There are many models of inflation, but the
single-field inflation in which a minimally coupled scalar field
(inflaton) is well motivated in the literatures. The action for these
single-field models leads to the equation of motion for the mode
functions \(u_{k}\) (or Mukhanov equation) \cite{Liddle:1993fq,Kundu:2011sg}:
\begin{equation}
	\label{f7}
u_{k}^{''} + \left( k^{2} - \frac{z^{''}}{z} \right)u_{k} = 0
\end{equation}
where $\frac{z^{''}}{z}=\frac{2c}{\tau^2}$, and $c=\frac{4\nu^{2}-1}{8}$. The most complete solutions of mode equation (\ref{f7}) can be written as below
in terms of Hankel functions \cite{Baumann:2009ds}:
\begin{equation}
	\label{f8}
u_{k} = \frac{\sqrt{\pi\tau}}{2}\left\lbrack A_{k}H_{\nu}^{(1)}\left( |k\tau| \right) + B_{k}H_{\nu}^{(2)}\left( |k\tau| \right) \right\rbrack
\end{equation}
where \(H_{\nu}^{(1)}\)and \(H_{\nu}^{(2)}\) are the first and second
kind of Hankel functions, respectively\cite{Armendariz-Picon:2003knj}. The power spectrum in
Hilbert space is defined as\cite{Liddle:1993fq}:
\begin{equation}
	\label{f9}
P_{\phi} = \frac{k^{3}}{2\pi^{2}}\left| u_{k}^{(P)} \right|^{2}
\end{equation}
By considering the background geometry method (\ref{f2}) and (\ref{f3}), and using a
similar relationship with (\ref{f5}), the following definition for the power
spectrum in Krein space can be expressed \cite{Mohsenzadeh:2014zfa}.
\begin{equation}
	\label{f10}
P_{\phi} = \frac{k^{3}}{2\pi^{2}}\left( \left| u_{k}^{(P)} \right|^{2} - \left| u_{k}^{(N)} \right|^{2} \right)
\end{equation}
where $P$ means solutions with positive frequency and related to the main
mode, and $N$ means solutions with negative frequency and related to the
background space-time. Also, the number of particles created in the
$k$ mode in Hilbert space is defined \cite{Mijic:1998if}:
\begin{equation}
	\label{f11}
\left\langle N \right\rangle = \frac{1}{4\omega_{k}(\tau)}\left| u_{k}^{'}(\tau) \right|^{2} + \frac{\omega_{k}(\tau)}{4}\left| u_{k}(\tau) \right|^{2} - \frac{1}{2}
\end{equation}
where $\omega_{k}^{2}(\tau)=k^{2} - \frac{z^{''}}{z}$. By considering the background geometry method (\ref{f2}) and (\ref{f3}), and using a
similar relationship with (\ref{f5}), the following definition for the spectrum
of particles created by purely geometric perturbations in Krein space
can be expressed \cite{Mohsenzadeh:2015mvz}.
\begin{equation}
	\label{f12}
\left\langle \Omega \middle| :N(\tau): \middle| \Omega \right\rangle = \left\langle \Omega \middle| N(\tau) \middle| \Omega \right\rangle - \left\langle 0 \middle| N(\tau) \middle| 0 \right\rangle
\end{equation}

where \(\left| \left. \ 0 \right\rangle \right.\ \) is the vacuum in the
background space-time and
\(\left| \left. \ \Omega \right\rangle \right.\ \) is the vacuum in the
curved space-time.
\section{Selection of Physical Geometry in Cosmology}
\label{Sec 4.}                         \subsection{The physical geometry is taken pure de Sitter}
In a pure de Sitter, the exact solution of the equation of motion (\ref{f7})
becomes \cite{Baumann:2009ds},
\begin{equation}
	\label{f13}
u_{k} = \frac{A_{k}}{a\sqrt{2k}}\left( 1 - \frac{i}{k\tau} \right)e^{- ik\tau} + \frac{B_{k}}{a\sqrt{2k}}\left( 1 + \frac{i}{k\tau} \right)e^{+ ik\tau}
\end{equation}
where $ A_{k}$ and $B_{k}$ are Bogoliubov coefficients. Commonly, this set of vacua is called $\alpha$-vacuum. The Bunch–Davies (BD) vacuum obtained by setting \(B_{k} = 0\) and
\(A_{k} = 1\) in the exact solution (\ref{f13}):
\begin{equation}
	\label{f14}
u_{k}^{BD} = \frac{e^{- ik\tau}}{a\sqrt{2k}}\left( 1 - \frac{i}{k\tau} \right)
\end{equation}
\begin{itemize}
	\item
	By adopting method (\ref{f2}), we choose the positive solution (\ref{f14}) from the
	pure de Sitter geometry and the negative solution from Minkowski
	geometry as background geometry \cite{Mohsenzadeh:2008yv}:
\end{itemize}
\begin{equation}
	\label{f15}
u_{k}^{ph} = u_{k}^{(P)} = \frac{e^{- ik\tau}}{a\sqrt{2k}}\left( 1 - \frac{i}{k\tau} \right),\quad u_{k}^{BG} = u_{k}^{(N)} = \frac{e^{+ ik\tau}}{a\sqrt{2k}}.
\end{equation}
Using (\ref{f10}) and (\ref{f15}), the power spectrum in Krein space can be obtained
\cite{Mohsenzadeh:2008yv,Sojasi:2010zz},
\begin{equation}
	\label{f16}
	 P_{\phi}(k) = \frac{H^{2}}{4\pi^{2}}
\end{equation}
which is the same as the standard result (scale invariant) \cite{Stewart:1993bc}.

\begin{itemize}
	\item
	By adopting method (\ref{f2}), and using (\ref{f12}) and (\ref{f15}), the spectrum of
	particles created by purely geometric perturbations in Krein space can
	be expressed \cite{Mohsenzadeh:2014zfa}.
\end{itemize}
\begin{equation}
	\label{f17}
\left\langle \Omega \middle| :N: \middle| \Omega \right\rangle = \frac{- 1}{2} + \frac{1}{4}|\frac{1}{\sigma_{0} ^{2}}-2|^{1/2}\big[\sigma_{0} +\sigma_{0}^{3} 
\big]+\frac{1}{4|\frac{1}{\sigma_{0} ^{2}}-2|^{1/2}}\big[\frac{1}{\sigma_{0}}-\sigma_{0}+\sigma_{0}^{3}
\big].
\end{equation}
Note that a similar result was obtained in \cite{Pereira:2009kv}.
\begin{itemize}
	\item
	By adopting method (\ref{f2}), we choose $\alpha$-vacua (\ref{f13}) as positive
	solution from the pure de Sitter geometry and the negative solution
	from Minkowski background geometry \cite{Sojasi:2012mz},
\end{itemize}
\begin{equation}
	\label{f18}
u_{k}^{ph} = u_{k}^{(P)} = \frac{A_{k}}{a\sqrt{2k}}\left( 1 - \frac{i}{k\tau} \right)e^{- ik\tau} + \frac{B_{k}}{a\sqrt{2k}}\left( 1 + \frac{i}{k\tau} \right)e^{+ ik\tau}
\end{equation}
and,
$$
u_{k}^{BG} = u_{k}^{(N)} = \frac{1}{a\sqrt{2k}}e^{+ ik\tau}
$$

Using (\ref{f10}) and (\ref{f18}), the power spectrum in Krein space can be obtained
\cite{Sojasi:2012mz},
\begin{equation}
	\label{f19}
P_{\phi} = \left( \frac{H}{2\pi} \right)^{2}\left\lbrack 1 - \sigma_{0}\sin\left( \frac{2}{\sigma_{0}} \right) \right\rbrack
\end{equation}
The value of
$\sigma_{0}=\frac{H_0}{\Lambda}$
is dimensionless, where $H_0$ is the initial Hubble expansion rate during inflation, and $\Lambda$ is the energy scale, e.g., Planck scale. This is
similar to the result of Danielson\textquotesingle s work (scale
dependent) \cite{Danielsson:2002kx}.

\begin{itemize}
	\item
	By adopting method (\ref{f3}), we choose the positive solution
	from pure de Sitter geometry \cite{Mohsenzadeh:2008yv}:
\end{itemize}
\begin{equation}
\label{f20}
u_{k}^{ph} = u_{k}^{(P)} = \frac{e^{- ik\tau}}{a\sqrt{2k}}\left( 1 - \frac{i}{k\tau} \right)
\end{equation}
also the background solution as with negative frequency as,
$$
u_{k}^{BG} = u_{k}^{(N)} = \frac{e^{+ ik\tau}}{a\sqrt{2k}}\left( 1 + \frac{i}{k\tau} \right).
$$
Using (\ref{f10}) and (\ref{f20}), the power spectrum in Krein space can be obtained
\cite{Mohsenzadeh:2008yv,Sojasi:2010zz,Mohsenzadeh:2011zz,Sojasi:2015sxa},
\begin{equation} 
	\label{f21}
	P_{\phi}(k) = \frac{H}{4\pi^{2}}ke^{- \alpha k^{2}}
\end{equation}
where \(\alpha = \frac{1}{\pi H^{2}}\) . Obviously, scale invariance breaks down here. This result is like to the work of others, where the scale invariance is broken due to consideration of gravitational field perturbations \cite{Kaloper:2003nv}.

\subsection{The physical geometry is taken approximate de Sitter}

As we know, the period of inflation begins in the approximate de Sitter space-time. Commonly, it is difficult to find a proper mode function in this high-energy region of the very early universe, where $H$ is variable. Therefore, we present the excited-de Sitter mode function as the fundamental mode during the inflation, so that it asymptotically approaches de Sitter mode function. The approximate mode can be obtained by expanding the Hankel function in (\ref{f8}) up to its third term \cite{Armendariz-Picon:2003knj}. Theretofore, we have  \cite{Mohsenzadeh:2013bba,Sojasi:2015rrn}:
\begin{equation}
	\label{f22}
u_{k} \cong \frac{e^{- ik\tau}}{a\sqrt{2k}}\left( 1 - \frac{i}{k\tau} - \frac{1}{2}\left( \frac{i}{k\tau} \right)^{2} \right)
\end{equation}
\begin{itemize}
	\item
	By adopting method (\ref{f3}), we choose (\ref{f22}) as positive solution and the
	negative solution from pure de Sitter geometry \cite{Mohsenzadeh:2013bba},
\end{itemize}
\begin{equation}
	\label{f23}
u_{k}^{ph} = u_{k}^{(P)} = \frac{e^{- ik\tau}}{a\sqrt{2k}}\left( 1 - \frac{i}{k\tau} - \frac{1}{2}\left( \frac{i}{k\tau} \right)^{2} \right)
\end{equation}
and,
$$
u_{k}^{BG} = u_{k}^{(N)} = \frac{e^{+ ik\tau}}{a\sqrt{2k}}\left( 1 + \frac{i}{k\tau} \right)
$$
Using (\ref{f10}) and (\ref{f23}), the power spectrum in Krein space can be
obtained \cite{Mohsenzadeh:2013bba,Mohsenzadeh:2015mvz},
\begin{equation}
	\label{f24}
P_{\phi} = \left( \frac{H}{2\pi} \right)^{2}\left\lbrack 1 + \frac{1}{4}\left( \sigma_{0} \right)^{2} \right\rbrack
\end{equation}
which is scale dependent and the correction is second order
of  $\sigma_{0}$. Note that in \cite{Kempf:2001fa,Kaloper:2002uj}, a
similar correction has been obtained.

\begin{itemize}
	\item
	By adopting method (\ref{f3}), and using (\ref{f12}) and (\ref{f23}), the spectrum of
	particles created by purely geometric perturbations in Krein space can
	be expressed \cite{Mohsenzadeh:2014zfa}.
\end{itemize}

\begin{equation}
	\langle \Omega|:N:|\Omega \rangle= 	\frac{1}{4}|\frac{1}{\sigma_{0} ^{2}}-2|^{1/2}\big[\sigma_{0}^{3} +\frac{\sigma_{0}^{5} }{4}
	\big]+\frac{1}{4|\frac{1}{\sigma_{0} ^{2}}-2|^{1/2}}\big[\sigma_{0}+\frac{5}{4\sigma_{0}^{3}}+\sigma_{0}^{5}
	\big]
\end{equation}

Note that in \cite{Pereira:2009kv}, in the context of \emph{f}(\emph{R}) gravity,
similar behavior has been discussed.

\begin{itemize}
	\item
	By adopting method (\ref{f3}), we choose the following solution as the
	general approximate solution, include positive and negative frequency,
	for approximate-dS space--time \cite{Yusofi:2014mma}:
\end{itemize}
\begin{equation}
	\label{f26}
u_{k} = \frac{A_{k}}{\sqrt{2k}}\left( 1 - \frac{i}{k\tau} - \frac{1}{2}\left( \frac{i}{k\tau} \right)^{2} \right)e^{- ik\tau}
+\frac{B_{k}}{\sqrt{2k}}\left( 1 + \frac{i}{k\tau} - \frac{1}{2}\left( \frac{i}{k\tau} \right)^{2} \right)e^{ik\tau}.
\end{equation}

where \(A_{k}\) and \(B_{k}\) are Bogoliubov coefficients. By adopting
the de Sitter geometry as background geometry and Similar to Danielsson
work \cite{Pereira:2009kv}, if we consider \(g_{k}\), corresponding to the first
derivative or conjugate momentum of \(u_{k}\) , we will have:
\begin{equation}
	\label{f27}
	g_{k} = A_{k}\sqrt{\frac{k}{2}}{\left( 1 - \frac{i}{k\tau} \right)}e^{-ik\tau} - B_{k}\sqrt{\frac{k}{2}}{\left( 1 + \frac{i}{k\tau} \right)}e^{ik\tau}
\end{equation}
Using (\ref{f9}), (\ref{f26}), and (\ref{f27}), the power spectrum in Hilbert space can be
obtained \cite{Yusofi:2014mma},
\begin{equation}
	\label{f28}
P_{\phi} \cong \left( \frac{H}{2\pi} \right)^{2}\left\lbrack 1 - \frac{i}{2}\left( \sigma_{0} \right)^{2}\cos\left( \frac{2}{\sigma_{0}} \right) \right\rbrack
\end{equation}
which is similar to the result of Martin\textquotesingle s work
\cite{Martin:2003kp}.

\begin{itemize}
	\item
	By adopting method (\ref{f2}), we choose the following solution as the
	general approximate solution for approximate-dS space--time (\ref{f26}) and by adopting the Minkowski geometry as background geometry and using the
Danielson method \cite{Pereira:2009kv}, we choose \(g_{k}\) ,corresponding to the
first derivative or conjugate momentum of \(u_{k}\) , we will have:\end{itemize}
\begin{equation}
	\label{f29}
g_{k} = A_{k}\sqrt{\frac{k}{2}}e^{- ik\tau} - B_{k}\sqrt{\frac{k}{2}}e^{ik\tau}
\end{equation}

Using (\ref{f9}), (\ref{f26}), and (\ref{f29}), the power spectrum in Hilbert space can be
obtained \cite{Yusofi:2014mma},
\begin{equation}
	\label{f30}
P_{\phi} \cong \left( \frac{H}{2\pi} \right)^{2}\left\lbrack 1 - \sigma_{0}\sin\left( \frac{2}{\sigma_{0}} \right)+\frac{1}{2}\left( \sigma_{0} \right)^{2}- \frac{1}{4}\left( \sigma_{0} \right)^{3}\sin\left(\frac{2}{\sigma_{0}} \right)+... \right\rbrack
\end{equation}

which is similar to the result of Martin\textquotesingle s work
\cite{Martin:2003kp}.

\subsection{The physical geometry is taken asymptotic de Sitter}

In mode equation (\ref{f7}), the function $z$ depend on the dynamics of the background space-time and is a time-dependent parameter. So, finding the exact solutions of (\ref{f7}) is difficult, therefore, the approximation or asymptotic methods are commonly used. In the very early universe, we have used asymptotic expansions of Hankel function up to the higher-order of  for the far past time limit   (\(|k\tau| \gg 1\)),   \cite{Armendariz-Picon:2003knj,Yusofi:2014pva}
$$
H_{\nu}^{(1,2)}\left( |k\tau| \right) \rightarrow \sqrt{\frac{2}{\pi|k\tau|}}\left\lbrack 1 \pm i\frac{4\nu^{2} - 1}{8|k\tau|} - \frac{\left( 4\nu^{2} - 1 \right)\left( 4\nu^{2} - 9 \right)}{2!\left( 8|k\tau| \right)^{2}} \pm \ldots \right\rbrack 
$$
\begin{equation}
	\label{f31}
\times \exp\left[\pm\left(|k\tau|-(\nu+\frac{1}{2})\pi\right)\right]
\end{equation}
according to the asymptotic expansion (\ref{f31}) and the equation of motion  (\ref{f7}), the most general form of mode functions
\(u_{k}\) for curved space-time becomes as \cite{Yusofi:2014mta,Yusofi:2014pva,Yusofi:2014vca,Yusofi:2015nna,Yusofi:2018udu}:
\begin{equation}
	\label{f32}
u_{k}(\tau,\nu) = A_{k}\frac{e^{- ik\tau}}{a\sqrt{2k}}\left( 1 - i\frac{c}{k\tau} - \frac{d}{k^{2}\tau^{2}} - \ldots \right)+ B_{k}\frac{e^{+ ik\tau}}{a\sqrt{2k}}\left( 1 + i\frac{c}{k\tau} - \frac{d}{k^{2}\tau^{2}} + \ldots \right)
\end{equation}

Where \(d = \frac{c(c - 1)}{2}\) , and \(A_{k}\) and \(B_{k}\) are
Bogoliubov coefficients. The positive frequency solutions of the mode
equation (\ref{f7}) become,
\begin{equation}
	\label{f33}
u_{k}^{(\nu)} = \frac{e^{- ik\tau}}{a\sqrt{2k}}\left\lbrack 1 - i\frac{c}{k\tau} - \frac{d}{k^{2}\tau^{2}} - \ldots \right\rbrack
\end{equation}

\begin{itemize}
	\item
	By adopting method (\ref{f2}), we choose the solution (\ref{f32}) as the general
	asymptotic solution, include positive and negative frequency, for
	asymptotic de Sitter space-time. Also, we adopt the Minkowski geometry as
	background geometry and similar to Danielsson work \cite{Danielsson:2002kx}, we
	consider (\ref{f29}) for \(g_{k}\) \cite{Mohsenzadeh:2016pxd}. Using (\ref{f9}) and (\ref{f32}), the power spectrum in Hilbert space can be obtained
\cite{Mohsenzadeh:2016pxd},\end{itemize}
\begin{equation}
	\label{f34}
P_{\phi}(k) = \left( \frac{H}{2\pi} \right)^{2}\left( \frac{2}{1 - \sqrt{8c + 1}} \right)^{2}\left\lbrack 1 - c\sigma_{0}\sin\left( \frac{2}{\sigma_{0}} \right)+\frac{5}{4}\left( c\sigma_{0} \right)^{2}- \left( c\sigma_{0} \right)^{3}\sin\left(\frac{2}{\sigma_{0}} \right)+... \right\rbrack
\end{equation}
Martin and Brandenberger obtained the corrections in \cite{Martin:2003kp}, such as we obtain in (\ref{f34}).

\begin{itemize}
	\item
	By adopting method (\ref{f3}), we choose the solution (\ref{f33}) as the general
	asymptotic solution, include positive and negative frequency, for
	asymptotic-dS space--time. Also, we adopt the de Sitter geometry as
	background geometry and Similar to Danielsson work \cite{Danielsson:2002kx}, we consider (\ref{f27}) for \(g_{k}\). 
Using (\ref{f9}), the power spectrum in Hilbert space can be obtained
 \cite{Mohsenzadeh:2016pxd}, \end{itemize}
\begin{equation}
	\label{f35}
P_{\phi}(k) = \left( \frac{H}{\pi\left( 1 - \sqrt{8c + 1} \right)} \right)^{2}\left\lbrack \frac{1}{1 - (c - 1)^{2}\sigma_{0}^{2}} \right\rbrack \left\lbrack 1 - (c-1)\sigma_{0}\sin\left( \frac{2}{\sigma_{0}} \right)+\frac{ (c-1)^{2}\sigma_{0} ^{2}}{4}\right\rbrack
\end{equation}
\begin{itemize}
	\item
	By adopting method (\ref{f3}), and using (\ref{f11}) and (\ref{f33}), the spectrum of
	particles created by purely geometric perturbations in Hilbert space
	can be expressed \cite{Yusofi:2018udu,Heydarzadeh:2021uht}.
\end{itemize}
\begin{equation}
	\label{f38}
\left\langle N \right\rangle_{gen} = \frac{\sigma_{0} ^{4}[d^{2} + (c - d)^{2}]}{4}
\end{equation}
the similar result previously was argued in \cite{Pereira:2009kv}.

\begin{itemize}
	\item
	By adopting method (\ref{f3}), and using (\ref{f9}) and (\ref{f33}), the tensor mode
	\(P_{h} = \frac{2}{M_{pl}}P_{\phi}\) in Hilbert space can be expressed
	\cite{Sojasi:2015rrn}.
\end{itemize}
\begin{equation}
	\label{f39}
P_{h} = \frac{H^{2}}{2\pi^{2}M_{pl}^{2}}\left\lbrack c + {\sigma_{0}}^{2}d^{2}\left( \frac{k}{k_{\circ}} \right)^{- \xi} \right\rbrack,
\end{equation}
where \(\xi\) is a free parameter. we choose \(k_{\circ}\) as a pivot scale, which is the largest observable scales initially exited the horizon. Our
results is consistent with the result of Planck team in 2018 data
\cite{Planck:2018jri}.

\begin{itemize}
	\item
	By adopting method (\ref{f3}), and using (\ref{f12}) and (\ref{f33}), the spectrum of
	particles created by purely geometric perturbations in Krein space can
	be expressed  \cite{Ziyaee:2020wik,Ziyaee:2021pno}.
\end{itemize}

\begin{equation}
	\label{f40}
	 \langle N \rangle_{phy}=\langle N \rangle_{adS}- \langle N \rangle_{dS}
\end{equation}
where
\begin{equation}
	\label{f41}
	\langle N \rangle_{adS}=\frac{1}{4}|\frac{1}{\sigma_{0} ^{2}}-2c|^{1/2}\big[\sigma_{0} +c\sigma_{0}^{3} 
	+d^{2}\sigma_{0}^{5}+...\big]+\frac{1}{4|\frac{1}{\sigma_{0} ^{2}}-2c|^{1/2}}\big[\frac{1}{\sigma_{0}}-c\sigma_{0}+(c-d)^{2}\sigma_{0}^{3}
	+4d^{2}\sigma_{0}^{5}+...\big],
\end{equation} 
and
\begin{equation}
	\label{f42}
\langle N \rangle_{adS}=\frac{1}{4}|\frac{1}{\sigma_{0} ^{2}}-2|^{1/2}\big[\sigma_{0} +\sigma_{0}^{3} 
\big]+\frac{1}{4|\frac{1}{\sigma_{0} ^{2}}-2|^{1/2}}\big[\frac{1}{\sigma_{0}}-\sigma_{0}+\sigma_{0}^{3}
\big]
\end{equation}
Note that the results of Krein method are the identical to the Hilbert (standard) method, when the background metric is flat.
\section{conclusions}
\label{Sec 5.}
In this paper, we first defined the quasi-de Sitter geometry and Krein space. Next, we examine the calculation of the power spectrum and the spectrum of particles created in this geometry by considering the definition of Krein space. The results of these calculations shows that, first of all, consideration of method (\ref{f3}) guarantees the preservation of the symmetry, which is a strict requirement for a high-energy universe. Also, the obtained spectra are without divergence and have automatically renormalized. An important feature of these spectrums is that they are scale dependent. This scale dependence can be due to gravitational field perturbations and graviton fluctuations in the early universe. In fact, these perturbations are caused due to considering pure de Sitter as the background geometry. In other words, considering the negative norm states in curved background geometry cause a spectrum that can have gravitational effects. Taking into account the gravitational effects in the early curved universe seems more logical than to ignore it. Because the early universe is the only place where we seek the unity and symmetry of forces, and also we are looking for quantum gravity. Therefore, according to the above justifications, the use of method (\ref{f3}) is more justifiable and defensible.\\
	\section*{Declaration of Competing Interest}
	The authors declare that they have no known competing financial interests or personal relationships that could have appeared to influence the work reported in this paper.
	\section*{Acknowledgements}
	This work has been supported by the Islamic Azad University, Qom Branch, Qom, Iran.\\
\bibliography{NNA_MOH_1402.bib} 

\end{document}